\documentclass[paper]{JHEP3} 

\usepackage{epsfig,multicol,multirow}
\usepackage{amsmath,subfigure,latexsym,amssymb,cite}

\newcommand\fverb{\setbox\fverbbox=\hbox\bgroup\verb}
\newcommand\fverbdo{\egroup\medskip\noindent%
			\fbox{\unhbox\fverbbox}\ }
\newcommand\fverbit{\egroup\item[\fbox{\unhbox\fverbbox}]}
\newbox\fverbbox

\newcommand {\beq} {\begin{equation}}
\newcommand {\eeq} {\end{equation}}
\newcommand {\beqa}{\begin{eqnarray}}
\newcommand {\eeqa}{\end{eqnarray}}

\newcommand{\1}{\mbox{1}\hspace{-0.25em}\mbox{l}}

\allowdisplaybreaks

\title{Realizing chiral fermions in the type IIB matrix model
at finite $N$
}

\author{
Jun Nishimura${}^{ab}$ and Asato Tsuchiya${}^{c}$
\vspace*{0.5cm} \\
\llap{$^a$}High Energy Accelerator Research Organization (KEK),\\
Tsukuba, Ibaraki 305-0801, Japan\\
\llap{$^b$}Department of Particle and Nuclear Physics,\\
Graduate University for Advanced Studies (SOKENDAI),\\
Tsukuba, Ibaraki 305-0801, Japan\\
\llap{$^c$}Department of Physics, Shizuoka University,\\
836 Ohya, Suruga-ku, Shizuoka 422-8529, Japan
\vspace*{0.5cm} \\
\email{jnishi@post.kek.jp, satsuch@ipc.shizuoka.ac.jp}}

\preprint{KEK-TH-1635}

\abstract{
We discuss how chiral fermions can appear in
the type IIB matrix model, which is considered to be 
a nonperturbative formulation of superstring theory.
In particular, we are concerned with 
a constructive definition of the theory,
in which we start with a finite-$N$ configuration
and take the large-$N$ limit later on.
We point out that there exists a certain necessary condition
which the structure of the extra dimensions should satisfy.
As an example, we consider a previous proposal using intersecting branes
and show that chiral fermions can indeed be realized
in four dimensions
by introducing a matrix counterpart of warped space-time.
This is remarkable in view of the well-known difficulty in 
realizing chiral fermions in lattice gauge theory. 
}

\keywords{Matrix Models, Superstring Vacua}

\begin{document}

\section{Introduction}
An important feature of the Standard Model of 
elementary particles
is that fermions are chiral.
It is quite nontrivial, however,
to get chiral fermions
from 
a fundamental theory 
in higher dimensions such as superstring theory
or from phenomenological models with extra dimensions.
Note, for instance, that fermions in higher dimensions
become vector-like in four dimensions if one applies
simple dimensional reduction.
A popular idea among many others is to use
orbifolding, which amounts
to imposing nontrivial 
boundary conditions in the extra dimensions.

In this paper we discuss how chiral fermions can appear
in the type IIB matrix model \cite{IKKT}, which is considered to
be a nonperturbative definition of superstring theory.
This is highly motivated, in particular, in view of 
recent developments in the Lorentzian version of the model.
In Ref.~\cite{Kim:2011cr}, Monte Carlo studies were made possible,
and it was shown that a well-defined large-$N$ limit can be taken
despite the fact that the action is not positive definite.
Moreover, the time evolution of the space was extracted
from the ten bosonic matrices,
and it was shown that 
three out of nine directions start to expand after some critical time,
signaling spontaneous breaking of nine-dimensional rotational symmetry.
Classical equations of motion are expected to be
valid at late times \cite{Kim:2011ts,Kim:2012mw}, 
and interesting solutions with
expanding behavior, which naturally solve the cosmological 
constant problem, were found \cite{Kim:2012mw}.
Ref.~\cite{Nishimura:2012rs} discussed
how one can obtain a local field theory from fluctuations around 
a classical solution representing a commutative space-time.

The issue of realizing chiral fermions in the type IIB matrix model
has been discussed by many authors.
For instance, Ref.~\cite{HajimeAoki} addressed this issue
by considering an ``effective theory'' for toroidally compactified
extra dimensions. The space in extra dimensions is represented
by six unitary matrices, and the overlap Dirac operator
is used to respect chirality.
Chiral fermions are indeed realized 
in the large-$N$ limit of
finite-$N$
classical backgrounds with nonvanishing fluxes in extra dimensions.
However, the connection between this ``effective theory'' and the
original model is not very clear.

More recently, Ref.~\cite{Chatzistavrakidis:2011gs} proposed
to realize chiral fermions in four dimensions
using matrix configurations representing intersecting branes.
This proposal is based on the original model unlike the one mentioned above.
%
Here we are concerned with
a constructive definition, in which
we start with a finite-$N$ configuration
and take the large-$N$ limit later on.
By explicit calculations, we confirm below that a chiral zero mode
of the Dirac operator indeed appears on the intersection
of the branes.
However, we also find that another chiral zero mode with opposite
chirality appears.
Because of this unwanted zero mode, we obtain a vector-like 
theory in four dimensions
if space-time is a direct product of our
four-dimensional space-time and the extra dimensions
as the authors implicitly assume.

In fact the above conclusion can be understood as 
a consequence of 
a ``no-go theorem'' that
applies to the original model, which is written in terms of
Hermitian matrices and with an ordinary Dirac operator.
This no-go theorem can be circumvented, however, 
by considering a matrix counterpart of the warped space-time,
which 
appears generically
in the type IIB matrix model.
We show that it is indeed possible to realize
chiral fermions in four dimensions.
This is remarkable
in view of the well-known difficulty in realizing chiral fermions
in lattice gauge theory.
While we need to show eventually that
a background that gives rise to 
chiral fermions
appears dynamically, 
we consider it very promising that such a finite-$N$ 
configuration
does exist in the type IIB matrix model.
%
%

The rest of this article is organized as follows.
In section \ref{sec:Lorentz-inv}
we discuss Lorentz invariant backgrounds and show
that a matrix version of the warp factor appears
in general.
In section \ref{sec:mode-expansion}
we expand the fermionic variables in terms of 
modes in the extra dimension.
In particular, we show that one can obtain 
a chiral fermion in 4d if 
the matrix version of the warp factor
satisfies certain conditions.
In section \ref{sec:example} we present an explicit
example of the background configurations that
give rise to a chiral fermion in 4d.
In section \ref{sec:gauge-symmetry} we show
how one can obtain non-Abelian gauge groups
in the present setup, and discuss some issues 
related to gauge interactions.
Section \ref{sec:summary} is devoted to a summary
and discussions.

\section{Lorentz invariant backgrounds}
\label{sec:Lorentz-inv}
In the type IIB matrix model, the space-time is represented
by the ten bosonic $N \times N$ Hermitian matrices $A_M$
($M = 0 , \ldots , 9$).
As it was shown by Ref.~\cite{Kim:2011cr}, an
expanding three-dimensional space appears dynamically
after some time.
At later times, it is speculated that three-dimensional space
becomes much larger than the Planck scale, and that 
quantum fluctuations can be neglected 
at large scales \cite{Kim:2011ts,Kim:2012mw}.
Furthermore, as far as we do not consider too long time scale,
we may neglect the expansion of space and therefore the space-time
has SO(3,1) 
Lorentz symmetry.
Thus we are led to consider matrix configurations given by
\begin{align}
A_\mu &= X_\mu \otimes M  \quad (\mu=0, \ldots ,3) \ ,
\label{Amu}
\\
A_a &= \1_n \otimes  Y_a  \quad (a=4, \ldots ,9) \ .
\label{Aa}
\end{align}
Here we assume that the $n \times n$ Hermitian matrices
$X_\mu$ have the property
$O_{\mu\nu} X_\nu = g[O] \, X_\mu \, g[O]^\dag$,
where $O \in {\rm SO}(3,1)$ and $g[O]\in {\rm SU}(n)$.
Then (\ref{Amu}), (\ref{Aa}) may be
regarded as the most general configuration that is
SO(3,1) 
invariant up to ${\rm SU}(N)$ symmetry.

The Hermitian matrix $M$ in (\ref{Amu})
may be regarded as a matrix version
of the warp factor.\footnote{This matrix $M$ introduces
noncommutativity such as $[A_\mu, A_a]\neq 0$ in general. 
However, as one can see from eq.~(\ref{chiral-S-f}), 
the noncommutativity does not show up in (3+1)-dimensional
field theory as far as massless modes are concerned.}
The special case $M=\1$ corresponds
to a space-time which is a direct product of (3+1)-dimensional
space-time and the extra dimensions.
However, from the viewpoint of the Lorentz symmetry, there is no reason
to set $M=\1$.
%

\section{Mode expansion of fermionic variables}
\label{sec:mode-expansion}

The type IIB matrix model has an action with the fermionic part
given by
\begin{align}
S_{\rm f}=\frac{1}{2}\, \mbox{Tr}
\Bigl(\bar{\Psi}\Gamma^M[A_M,\Psi] \Bigr) \ ,
\label{fermionic action}
\end{align}
where $\Gamma_M$ are $32 \times 32$ gamma matrices
in 10d.
The fermionic matrices $\Psi _\alpha$ ($\alpha = 1, \ldots , 32$)
are Majorana-Weyl fermions in 10d,
and, in particular, they satisfy
\begin{align}
\Gamma_{\chi} \Psi  &= \Psi \ ,
\label{Weyl-cond}
\end{align}
where $\Gamma_{\chi}$ is
the chirality operator in 10d.

In order to discuss chiral fermions in 4d, 
it is convenient to decompose the gamma matrices in 10d 
into the ones in 4d and 6d as
\begin{align}
\Gamma^{\mu}&=\gamma^{\mu}\otimes \1_8 \ , \nonumber\\
\Gamma^a&= i \gamma_{\chi}^{\rm (4d)} \otimes \Delta^a \ ,
\label{gamma matrices}
\end{align}
where 
$\gamma^{\mu}$ and $\Delta^a$ are gamma matrices in 4d and 6d, 
respectively, which satisfy
\begin{align}
\{\gamma^{\mu},\gamma^{\nu}\}&=-2\eta^{\mu\nu} \ , \nonumber\\
\{\Delta^a,\Delta^b\}&=2\delta^{ab} \ ,
\end{align}
and $\gamma_{\chi}^{(4d)}$ is the chirality operator in 4d.
Note that the chirality operator
$\Gamma_{\chi}$ in 10d can be decomposed as
\begin{align}
\Gamma_{\chi} = \gamma_{\chi}^{\rm (4d)} \otimes
\Delta_{\chi}^{\rm (6d)} \ ,
\label{chi-decompose}
\end{align}
where $\Delta_{\chi}^{\rm (6d)}$ is
the chirality operator in 6d.

In the case of quantum field theory in higher dimensions,
one decomposes fields into
Kaluza-Klein modes, which can then be identified
as four-dimensional fields.
Here we make a similar analysis in the language of 
matrices.\footnote{The analysis presented here is improved
from the one given in the first version of this paper, in which
we discussed the solutions to the Dirac equation in 10d.
In particular, the improved analysis is free from the problem
pointed out in the footnote 1 of Ref.~\cite{Steinacker:2013eya}.}
We consider expanding the fermionic variables
in terms of the eigenmodes of
the Dirac operator in 6d defined by
\begin{align}
D_{\rm 6d}\Phi=\Delta^a[Y_a,\Phi] \ .
\label{Dirac operator in 6d}
\end{align}
In the explicit example to be discussed in the next section,
we consider a configuration of $Y_a$, which has a block diagonal form
\begin{align}
Y_a= \left( 
      \begin{array}{cc}
       Y_a^{(1)} & 0 \\
       0   & Y_a^{(2)} 
      \end{array} \right) \ .
\label{Ya-block}
\end{align}
Correspondingly, we decompose $\Phi$ 
in eq.(\ref{Dirac operator in 6d}) as
\begin{align}
\Phi= \left( 
      \begin{array}{cc}
       \Phi^{(1,1)} & \Phi^{(1,2)} \\
       \Phi^{(2,1)}   & \Phi^{(2,2)}
      \end{array} \right) \ .
\label{decomposition into blocks}
\end{align}
Since the Dirac operator $D_{\rm 6d}$ acts 
on each block $\Phi^{(I,J)} \;(I,J=1,2)$ independently,
the eigenvalue problem for $D_{\rm 6d}$ can be decomposed
into that in each block.

As we will see in the explicit example,
chiral fermions actually appear in off-diagonal blocks.
Therefore, from now on,
we consider the eigenvalue problem for 
$\varphi \equiv \Phi^{(1,2)}$, which is given by
\begin{align}
\Delta^a(Y_a^{(1)}\varphi-\varphi Y_a^{(2)})=\lambda \, \varphi \ .
\label{eigenvalue equation}
\end{align}
Due to the fact that $Y_a$ and $\Delta^a$ are Hermitian matrices, 
one can easily show that the eigenvalue $\lambda$ in 
(\ref{eigenvalue equation}) is real.
Also, by multiplying $\Delta_{\chi}^{\rm (6d)}$ to
(\ref{eigenvalue equation}) from the left, one obtains
\begin{align}
\Delta^a \Bigl\{ Y_a^{(1)}( \Delta_{\chi}^{\rm (6d)}\varphi)
-(\Delta_{\chi}^{\rm (6d)}\varphi) Y_a^{(2)} \Bigr\}
=-\lambda \, (\Delta_{\chi}^{\rm (6d)}\varphi) \ .
\label{eigenvalue equation 2}
\end{align}
This implies that 
if $\varphi$ is an eigenvector with 
the eigenvalue $\lambda$, 
$\Delta_{\chi}^{\rm (6d)}\varphi$
is an eigenvector with the eigenvalue $-\lambda$.
In particular, $\varphi$ and $\Delta_{\chi}^{\rm (6d)}\varphi$ 
are linearly independent for $\lambda \neq 0$.
Therefore we can construct left-handed and right-handed modes 
by taking linear combinations of
$\varphi$ and $\Delta_{\chi}^{\rm (6d)}\varphi$ 
with $\lambda \neq 0$ as
\begin{align}
\varphi_R &=\frac{1+\Delta_{\chi}^{\rm (6d)}}{2}\varphi \ , \\
\varphi_L &=\frac{1-\Delta_{\chi}^{\rm (6d)}}{2}\varphi \ ,
\end{align}
which satisfy
\begin{align}
\Delta_{\chi}^{\rm (6d)}\varphi_R&=\varphi_R  \ , \nonumber\\
\Delta_{\chi}^{\rm (6d)}\varphi_L&=-\varphi_L  \ , 
\label{6d chirality}
\end{align}
and
\begin{align}
\Delta^a(Y_a^{(1)}\varphi_R-\varphi_R Y_a^{(2)})
&=\lambda \, \varphi_L \ , \nonumber\\
\Delta^a(Y_a^{(1)}\varphi_L-\varphi_L Y_a^{(2)})
&=\lambda \, \varphi_R \ .
\label{6d Dirac equation}
\end{align}
Thus the non-zero modes appear in pairs of right-handed and left-handed modes.
On the other hand,
the zero modes can be 
assumed to have definite chirality.
Since we are considering finite-$N$ matrices,
the space of $\varphi$ with each chirality
has the same dimension.
Therefore, the number of zero modes with each chirality 
should also be the same.
However, the actual form of the zero mode $\varphi$ with each chirality
can be very different in general. 
This fact will be important in getting chiral fermions in 4d.

Let $\{\lambda_n\}$ be a set of non-negative eigenvalues 
in (\ref{eigenvalue equation}).
Then we denote the right-handed and left-handed modes 
corresponding to $\lambda_n$ 
by $\varphi_{nR}$ and $\varphi_{nL}$, respectively.
These modes can be normalized in such a way that 
they satisfy the orthonormal condition
\begin{align}
\mbox{tr}(\varphi_{m A}^{\dagger}\varphi_{n B})=\delta_{mn}\delta_{AB} \ ,
\label{orthonormality condition}
\end{align}
where $A$ and $B$ are $R$ or $L$.

Now we decompose the fermionic variables $\Psi$ in 
(\ref{fermionic action})
in the same way as in (\ref{decomposition into blocks}), and 
expand the off-diagonal block $\Psi^{(1,2)}$ 
in terms of the orthonormal basis 
$\varphi_{nR}$ and $\varphi_{nL}$ constructed above as
\begin{align}
\Psi^{(1,2)}=\sum_n (\psi_{nR}\otimes \varphi_{nR} 
+ \psi_{nL}\otimes \varphi_{nL} ) \ .
\label{expansion of Psi}
\end{align}
Note that the matrix coefficients $\psi_{nR}$ and $\psi_{nL}$
introduced here satisfy
\begin{align}
\gamma_{\chi}^{\rm (4d)} \psi_{nR} &= \psi_{nR}  \ , \nonumber\\
\gamma_{\chi}^{\rm (4d)} \psi_{nL} &= -\psi_{nL} \ ,
\label{4d chirality}
\end{align}
as one can see from (\ref{Weyl-cond}), (\ref{chi-decompose}) 
and (\ref{6d chirality}).
Namely, the left-handed and right-handed modes 
in 6d correspond to the left-handed and right-handed modes 
in 4d, respectively.
Note also that the other off-diagonal block $\Psi^{(2,1)}$
is related to the off-diagonal block $\Psi^{(1,2)}$ that
we have considered through charge conjugation as
\begin{align}
\Bigl(\Psi^{(2,1)}\Bigr)^{\rm c}=\Psi^{(1,2)}  \ ,
\label{Majorana condition}
\end{align}
due to the Majorana condition for $\Psi$.
This allows us to focus only on $\Psi^{(1,2)}$.

In what follows we consider the case in which
the matrix warp factor $M$ in (\ref{Amu}) has a block diagonal form
similar to (\ref{Ya-block}) as
\begin{align}
M =\left( \begin{array}{cc}
                M^{(1)} & 0 \\
                0 & M^{(2)}
        \end{array} \right) \ .
\label{block diagonal M}
\end{align}
Then, by substituting (\ref{expansion of Psi}) into the action 
(\ref{fermionic action}), we obtain
\begin{align}
S_{\rm f}
&=\sum_{m,n}(\mbox{tr}(\bar{\psi}_{mR}
\gamma^{\mu}X_{\mu}\psi_{nR})
\mbox{tr}(\varphi_{mR}^{\dagger}M^{(1)}\varphi_{nR})
-\mbox{tr}(\bar{\psi}_{mR}\gamma^{\mu}\psi_{nR}X_{\mu})
\mbox{tr}(\varphi_{mR}^{\dagger}\varphi_{nR}M^{(2)})) \nonumber\\
&\;\;\;+\sum_{m,n}(\mbox{tr}(\bar{\psi}_{mL}\gamma^{\mu}X_{\mu}\psi_{nL})
\mbox{tr}(\varphi_{mL}^{\dagger}M^{(1)}\varphi_{nL})
-\mbox{tr}(\bar{\psi}_{mL}\gamma^{\mu}\psi_{nL}X_{\mu})
\mbox{tr}(\varphi_{mL}^{\dagger}\varphi_{nL}M^{(2)})) \nonumber\\
&\;\;\;+i\sum_n\lambda_n\mbox{tr}(\bar{\psi}_{nR}\psi_{nL} 
-\bar{\psi}_{nL} \psi_{nR}) \nonumber\\
&\;\;\;+(\mbox{contribution from $\Psi^{(1,1)}$ and $\Psi^{(2,2)}$}) \ ,
\label{fermionic action 2}
\end{align}
where we have used (\ref{gamma matrices}), (\ref{6d Dirac equation}), 
(\ref{orthonormality condition}), (\ref{4d chirality}), 
(\ref{Majorana condition}) and (\ref{block diagonal M}).

Let us first consider the case $M=\1$, namely 
$M^{(1)}=\1$ and $M^{(2)}=\1$ in (\ref{block diagonal M}).
Then the action (\ref{fermionic action 2}) reduces to
\begin{align}
S_{\rm f}
&=\sum_n\mbox{tr}(\bar{\psi}_{nR}\gamma^{\mu}[X_{\mu},\psi_{nR}]
+\bar{\psi}_{nL}\gamma^{\mu}[X_{\mu},\psi_{nL}]
+i\lambda_n(\bar{\psi}_{nR}\psi_{nL} -\bar{\psi}_{nL} \psi_{nR}))
\nonumber  \\
&\;\;\; +(\mbox{contribution from $\Psi^{(1,1)}$ and $\Psi^{(2,2)}$}) \ ,
\end{align}
which implies that we obtain vector-like fermions in 4d in this case. 
(In section \ref{sec:gauge-symmetry}, we show
that the interaction with the gauge field is also vector-like.)
Obviously the same conclusion is reached
in the case of $Y_a=0$ with general $M$.
Therefore we find that in order to obtain chiral fermions,
we need to consider $M \neq \1$ and nonvanishing $Y_a$.
This is a necessary condition for the appearance of 
chiral fermions in 4d in the present setup.

As an example of configurations that give rise to chiral fermions in 4d,
we consider the case with 
$\lambda_0=0$ and $\lambda_n \neq 0$ for $n \neq 0$
and with $M^{(1)}$ and $M^{(2)}$ satisfying
\begin{align}
M^{(1)}\varphi_{0L}=\varphi_{0L}M^{(2)}=\varphi_{0L} \ .
\label{condition for M}
\end{align}
Then, the action (\ref{fermionic action 2}) reduces to
\begin{align}
S_{\rm f}
&=\mbox{tr}(\bar{\psi}_{0L}\gamma^{\mu}[X_{\mu},\psi_{0L}]) \nonumber\\
&\;\;\;
+\sum_{m}(\mbox{tr}(\bar{\psi}_{mR}\gamma^{\mu}X_{\mu}\psi_{0R})
\mbox{tr}(\varphi_{mR}^{\dagger}M^{(1)}\varphi_{0R})
-\mbox{tr}(\bar{\psi}_{mR}\gamma^{\mu}\psi_{0R}X_{\mu})
\mbox{tr}(\varphi_{mR}^{\dagger}\varphi_{0R}M^{(2)}))
+\mbox{c.c.} \nonumber\\
&\;\;\;
+\sum_{m,n\neq 0}(\mbox{tr}(\bar{\psi}_{mR}\gamma^{\mu}X_{\mu}\psi_{nR})
\mbox{tr}(\varphi_{mR}^{\dagger}M^{(1)}\varphi_{nR})
-\mbox{tr}(\bar{\psi}_{mR}\gamma^{\mu}\psi_{nR}X_{\mu})
\mbox{tr}(\varphi_{mR}^{\dagger}\varphi_{nR}M^{(2)})) \nonumber\\
&\;\;\;+\sum_{m,n\neq 0}
(\mbox{tr}(\bar{\psi}_{mL}\gamma^{\mu}X_{\mu}\psi_{nL})
\mbox{tr}(\varphi_{mL}^{\dagger}M^{(1)}\varphi_{nL})
-\mbox{tr}(\bar{\psi}_{mL}\gamma^{\mu}\psi_{nL}X_{\mu})
\mbox{tr}(\varphi_{mL}^{\dagger}\varphi_{nL}M^{(2)})) \nonumber\\
&\;\;\;+i\sum_{n\neq 0}\lambda_n
\mbox{tr}(\bar{\psi}_{nR}\psi_{nL} -\bar{\psi}_{nL} \psi_{nR}) \nonumber\\
&\;\;\;+(\mbox{contribution from $\Psi^{(1,1)}$ and $\Psi^{(2,2)}$}) \ .
\label{chiral-S-f}
\end{align}
Thus we find that
$\psi_{0L}$ has an appropriate form of the action as 
a chiral fermion in 4d, and
it does not couple to the other modes. 
On the other hand, $\psi_{0R}$ is coupled 
to the massive modes with $\lambda_n \neq 0$.
This implies that we obtain a left-handed chiral fermion in 4d
but not a right-handed chiral fermion.


\section{Explicit example}
\label{sec:example}

As an example of $Y_a$, which allows nontrivial zero eigenvalue
in (\ref{eigenvalue equation}), we consider a setup
representing intersecting branes proposed in
Ref.~\cite{Chatzistavrakidis:2011gs}.
Here we consider an explicit finite-$N$ configuration
given by
\begin{align}
Y_4 &= \frac{1}{r} \left(
\begin{array}{cc}
L_1 & 0 \\
0   & \1_k \otimes \tilde{L}_3
\end{array} \right) \ ,
&
Y_7 &= \frac{1}{r} \left(
\begin{array}{cc}
0  & 0 \\
0   & L_2 \otimes \1_k
\end{array} \right) \ ,
\nonumber \\
Y_5 &= \frac{1}{r} \left(
\begin{array}{cc}
L_2 & 0 \\
0   & 0
\end{array} \right) \ ,
&
Y_8 &=\frac{1}{r} \left(
\begin{array}{cc}
0   & 0 \\
0   & \tilde{L}_3 \otimes L_1
\end{array} \right) \ ,
\nonumber \\
Y_6 &= \frac{1}{r} \left(
\begin{array}{cc}
\tilde{L}_3 & 0 \\
0   & L_1 \otimes \1_k
\end{array} \right) \ ,
&
Y_9 &= \frac{1}{r} \left(
\begin{array}{cc}
0   & 0 \\
0   & \1_k \otimes L_2
\end{array} \right) \ ,
\label{CSZ-config}
\end{align}
where the $k\times k$ matrices $L_i$ ($i=1,2,3$)
are 
the $k$-dimensional irreducible representation of SU(2) algebra,
which represents a fuzzy sphere
$\sum_{i=1}^3 (L_i)^2 = r^2  \1 $
with the radius $r=\frac{1}{2} \sqrt{k^2 -1}$.
In Eq.~(\ref{CSZ-config})
we have also defined $\tilde{L}_3\equiv  L_3+ r \1_k$
so that the D5-branes represented
by the top-left block ($Y_a ^{(11)}$) and 
the D7-branes represented by
the bottom-right block ($Y_a ^{(22)}$) 
intersect once at the origin of 6d space.

\begin{figure}[htb]
\begin{center}
\includegraphics[height=5.5cm]{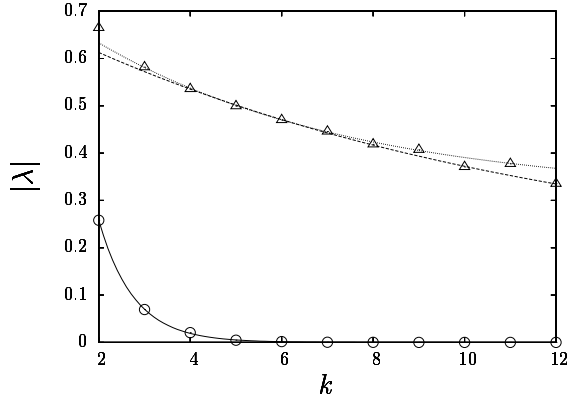}
\end{center}
\caption{
The smallest $|\lambda|$ (circles) and the second smallest one (triangles)
are plotted against $k$. The solid, dashed and dotted lines 
represent the fits to the behavior 
$|\lambda| = a + b \, {\rm exp}(- c \, k)$.
For the smallest $|\lambda|$, we find $a = 0.0001(4)$, which is consistent
with zero within the fitting error.
For the second smallest one, we find that the fitting curve
depends on the parity of $k$, and we obtain 
$a = 0.31(1)$ and $a = 0.15(3)$ for odd $k$ and even $k$,
respectively.
}
\label{fig:lambda-k}
\end{figure}

We solve Eq.~(\ref{eigenvalue equation}) for $\varphi$
with $Y_a$ given above.
Since 
$\varphi \equiv \Phi^{(1,2)}$
represents the off-diagonal block,
it actually corresponds to the degrees of freedom
connecting the two different branes.
In Fig.~\ref{fig:lambda-k} the smallest $|\lambda|$
and the second smallest one are plotted against $k$.
We find that the smallest one vanishes rapidly with increasing $k$,
which implies the appearance of two zero modes with each chirality in the large-$k$ limit.
The second smallest $|\lambda|$, on the other hand, 
seems to approach a non-zero constant, which depends on the parity of $k$.

\begin{figure}[htb]
\begin{center}
\includegraphics[height=4.0cm]{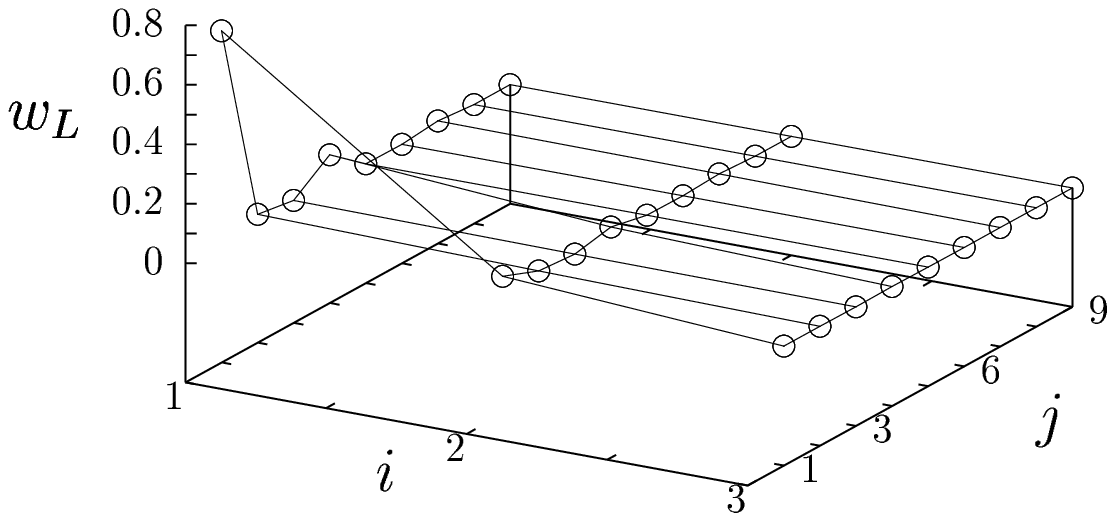}
\end{center}
\caption{
The shape of the wave function for the left-handed mode, $w_L(i,j)$.
A clear peak is seen at $(i,j)=(1,1)$
already in the $k=3$ case shown here.
}
\label{fig:wfl}
\end{figure}

\begin{figure}[htb]
\begin{center}
\includegraphics[height=4.0cm]{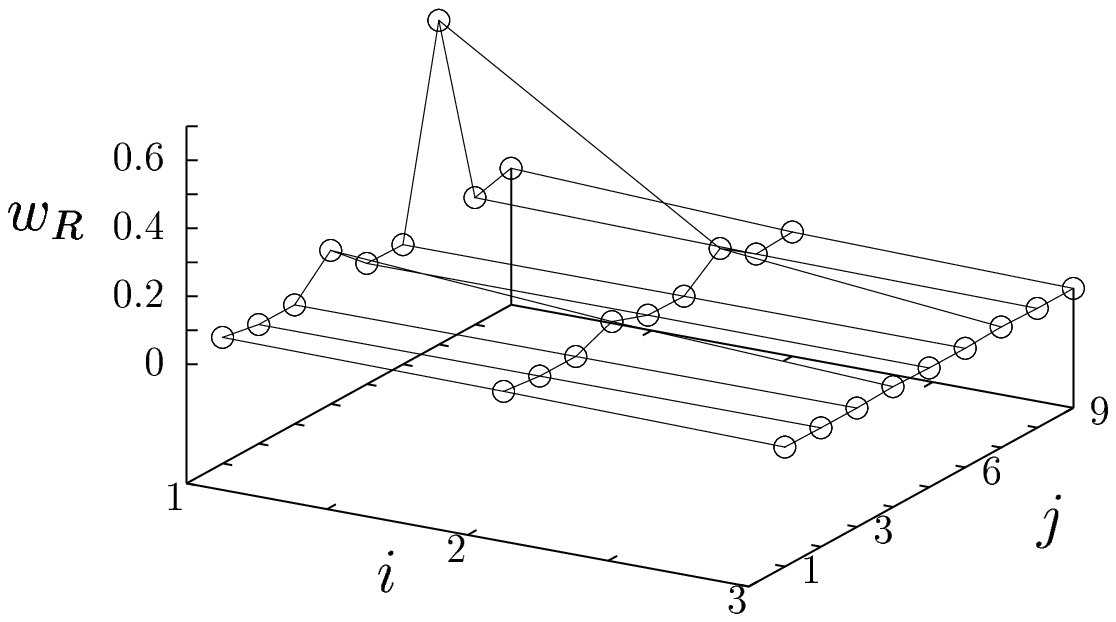}
\end{center}
\caption{
The shape of the wave function for the right-handed mode, $w_R(i,j)$, for $k=3$. 
A peak is seen at $(i,j)=(1,j_0)$, where $j_0=k(k-1)+1$.
}
\label{fig:wfr}
\end{figure}

We define the shape of the wave function 
for the left-handed and right-handed modes with the smallest
$\lambda$ by
\begin{align}
w_L(i,j) &\equiv \sum_{\alpha} |(\varphi_{L\alpha})_{ij}|^2 \ , 
\nonumber\\
w_R(i,j) &\equiv \sum_{\alpha} |(\varphi_{R\alpha})_{ij}|^2 \ ,
\end{align}
respectively. 
Figs.~\ref{fig:wfl} and  \ref{fig:wfr} shows $w_L$ and $w_R$, respectively. 
We see that $w_L$ has 
a peak at $(i,j)=(1,1)$, where $i=1$ and $j=1$ correspond
to the intersecting point on the two branes, while
$w_R$ has a 
peak at $(i,j)=(1,j_0)$, where $j_0=k(k-1)+1$ corresponds 
to a point on the second brane,
which is far from the intersecting point.
As the matrix size is increased, the peak of the left-handed mode
at $(i,j)=(1,1)$ becomes higher, and all the other components of $w_L(i,j)$ 
vanish in the large-$k$ limit.
On the other hand, the peak of the right-handed mode becomes lower 
as the matrix size is increased.

Due to the simple wave function for the left-handed mode,
it is easy to find the matrices $M^{(1)}$ and $M^{(2)}$ which satisfy 
(\ref{condition for M}) for this mode in the large-$k$ limit. 
They are 
given by (e.g., in the case of $k=3$)
\begin{align}
M^{(1)} &=\left(
\begin{array}{ccc}
1 & 0  & 0 \\
0 & *  & * \\
0 & *  & * 
\end{array} \right) \ , 
\nonumber\\
M^{(2)} &=\left(
\begin{array}{ccccccccc}
1 & 0  & 0 & 0 & 0 & 0 & 0 & 0 & 0\\
0 &&&&&&&&\\
0 &&&&&&&&\\
0 &&&&&&&&\\
0 &&&& * &&&&\\
0 &&&&&&&&\\
0 &&&&&&&&\\
0 &&&&&&&&\\
0 &&&&&&&&         
\end{array} \right) \ ,
\label{M-example}
\end{align}
because $\varphi_L$ takes the form 
\begin{align}
\varphi_{L\alpha}=\left(
\begin{array}{ccccccccc}
\chi_{\alpha} & 0  & 0 & 0 & 0 & 0 & 0 & 0 & 0\\
0             & 0  & 0 & 0 & 0 & 0 & 0 & 0 & 0\\
0             & 0  & 0 & 0 & 0 & 0 & 0 & 0 & 0    
\end{array} \right) \ ,
\end{align}
where $\sum_{\alpha}|\chi_{\alpha}|^2=1$.


\section{Gauge symmetry}
\label{sec:gauge-symmetry}
We can introduce gauge symmetry 
by replacing each of the branes by coincident multiple branes
similarly to the case of D-brane effective theory.
For instance, if we make a replacement
\begin{align}
Y_a ^{(1)}  \mapsto Y_a ^{(1)} \otimes \1_{p} \ , \quad
Y_a ^{(2)}  \mapsto Y_a ^{(2)} \otimes \1_{q}  
\label{multiple-branes}
\end{align}
in the configuration (\ref{CSZ-config}),
we obtain ${\rm U}(p) \times {\rm U}(q)$ gauge symmetry
as a subgroup of the U($N$) symmetry of the original model.
Then the chiral zero mode that appears from the top-right block of
$\varphi$ becomes a bifundamental representation $(p , \bar{q})$.
%

Let us discuss how the chiral zero modes interact with the gauge field. 
The gauge field is expected to appear from the fluctuation $a_\mu$
of $A_\mu$ around (\ref{Amu}).
Let us decompose $a_\mu$ into blocks as
we have done in eq.~(\ref{decomposition into blocks}).
Then the four-dimensional massless modes appear from
the diagonal blocks $a_\mu^{(I,I)} \ (I=1,2)$,
which can be expanded as 
$a_\mu^{(I,I)} = \sum_n \tilde{a}_{\mu n}^{(I)} 
\otimes b_{n}^{(I)}$
as in eq.~(\ref{expansion of Psi}). 
The gauge field $\tilde{a}_{\mu 0}^{(I)}$
corresponds to the zero mode $b_0^{(I)}$,
which satisfies
$[Y_a^{(I)} , [Y_a^{(I)}  , b_0^{(I)} ]] = 0 $
in the Lorentz gauge.
This results in 
$b_0^{(I)} \propto \1$. 
The gauge field is therefore insensitive to the wave function 
in the extra dimensions.
This implies, in particular, that in the $M=\1$ case the chiral zero
modes with opposite chirality that appear from the same block
interact with the gauge field in the same manner.
Thus we obtain a vector-like gauge theory in that case.


It is not clear how gauge anomaly cancellation can play any role
in the present construction.
We speculate that a local field theory is obtained at low energy 
only when the gauge group and the fermion
contents are chosen such that the gauge anomaly is canceled.
This important issue is left for future investigations.


\section{Summary and discussions}
\label{sec:summary}
In this paper we pointed out a necessary condition for
the appearance of chiral fermions in the type IIB matrix model.
With this new insight, we reconsidered the recent proposal 
based on the idea of intersecting branes.
By studying explicit finite-$N$ configurations,
we found that
unwanted chiral zero modes with opposite chirality
appear as anticipated from our general argument.
In spite of this problem, we found a simple way to remove the
unwanted chiral zero modes.
A crucial role was played by a matrix version of the warp factor,
which appears generically in the Lorentz invariant background.
We emphasize that this is the first explicit example in which
chiral fermions in four dimensions 
are obtained
from the type IIB matrix model
in the large-$N$ limit starting with finite-$N$ configurations.

An important property of the type IIB matrix model is that
it has a built-in mechanism that generates (3+1)-dimensional
space-time in the Lorentzian case \cite{Kim:2011cr}.
We consider it very encouraging that the model also allows the
appearance of chiral fermions in (3+1)-dimensions
as the explicit example demonstrates.
This gives us strong motivations to proceed to the next step
and to think how the background that allows chiral fermions 
can be generated dynamically.
While our discussion given in this paper
is not restricted to the Lorentzian case, 
the signature of the space-time plays an important role 
in actual dynamics.
On top of the determination of the space-time dimensionality,
the dominance of a classical configuration at late times is expected
in the Lorentzian case \cite{Kim:2011ts,Kim:2012mw}.
We therefore consider it important to extend the Monte Carlo
studies in Ref.~\cite{Kim:2011cr} to much later times and/or
to study the late time behaviors by classical equations
of motion \cite{Kim:2011ts,Kim:2012mw} with 
nontrivial structures \cite{Steinacker:2011wb} 
(and possibly with quantum corrections)
in the extra dimensions.

The ``warp factor'', which we find important
in realizing chiral fermions in the type IIB matrix model,
may also play important roles in solving 
the hierarchy problem \cite{Randall:1999ee} and in accounting
for the difference among three generations of fermions.
In this regard it would be interesting to calculate the Yukawa
couplings in a manner similar to the work 
in a closely related field theory \cite{Abe:2012fj}.

\acknowledgments

We thank H.~Aoki, S.-W.~Kim,
H.~Steinacker and J.~Zahn
for valuable discussions.
Computations were carried out
on SR16000 at YITP and FX10 at University of Tokyo.
This work is supported in part by Grant-in-Aid
for Scientific Research
(No.\ 20540286, 24540264, and 23244057)
from JSPS.

\end{document}